\documentclass[aps,preprint,prd,showpacs,nofootinbib]{revtex4}

\usepackage{amsmath,amssymb,amsfonts,array}
\usepackage{bm,bbm}
\usepackage{cases}
\usepackage{dcolumn}
\usepackage{graphicx}
\usepackage{hyperref}
\usepackage{subfigure}
\usepackage{wasysym}
\usepackage{xcolor}

\definecolor{dullred}{rgb}{0.706,0.208,0.192}
\definecolor{darkred}{rgb}{0.545,0,0}
\definecolor{MaroonC}{rgb}{0,0.502,0.502}
\definecolor{dullblue}{rgb}{0,0.298,0.49}
\definecolor{blue3}{RGB}{31, 119, 180}
\definecolor{dullpurple}{rgb}{0.431,0.188,0.534}
\definecolor{darkgreen}{rgb}{0.075,0.302,0.047}
\definecolor{darkergreen}{rgb}{0,0.196,0.125}
\definecolor{darkergreen2}{rgb}{0,0.294,0.188}

\hypersetup{colorlinks=true,
	breaklinks=true,
	pdfstartview=Fit,
	linkcolor=blue,
	citecolor=blue,
	urlcolor=blue}

\def\be{\begin{equation}}
\def\ee{\end{equation}}
\def\ba{\begin{eqnarray}}
\def\ea{\end{eqnarray}}

\def\blue{\color{blue}}

\def\nn{\nonumber}
\def\lf{\left}
\def\rt{\right}

\begin{document}
	

\title{Climbing over the potential barrier during inflation via null energy condition violation}

\author{Shi Pan$^{1,5,6}$\footnote{\texttt{\blue panshi22@mails.ucas.ac.cn} (Corresponding author)}}
\author{Yong Cai$^{2}$\footnote{\texttt{\blue caiyong@zzu.edu.cn} (Corresponding author)}}
\author{Yun-Song Piao$^{1,3,4,5}$\footnote{\texttt{\blue yspiao@ucas.ac.cn} (Corresponding author)}}

\affiliation{$^1$ School of Fundamental Physics and Mathematical Sciences, Hangzhou Institute for Advanced Study, UCAS, Hangzhou 310024, China}

\affiliation{$^2$ Institute for Astrophysics, School of Physics, Zhengzhou University, Zhengzhou, Henan 450001, China}

\affiliation{$^3$ School of Physics, University of Chinese Academy
of Sciences, Beijing 100049, China}

\affiliation{$^4$ International Center for Theoretical Physics
Asia-Pacific, Beijing/Hangzhou, China}

\affiliation{$^5$ Institute of Theoretical Physics, Chinese
Academy of Sciences, P.O. Box 2735, Beijing 100190, China}

\affiliation{$^6$  University of Chinese Academy of Sciences,
Beijing 100190, China}

\begin{abstract}

The violation of the null energy condition (NEC) may play a crucial role in enabling a scalar field to climb over high potential barriers, potentially significant in the very early universe. We propose a single-field model where the universe sequentially undergoes a first stage of slow-roll inflation, NEC violation, and a second stage of slow-roll
inflation. Through the NEC violation, the scalar field climbs over high potential barriers, leaving unique characteristics on the primordial gravitational wave power spectrum, including a blue-tilted nature in the middle-frequency range and diminishing oscillation amplitudes at higher frequencies. Additionally, the power spectrum exhibits nearly scale-invariant behavior on both large and small scales.

\end{abstract}

\maketitle

\section{Introduction}

The detection of gravitational waves (GWs) originating from binary pulsar systems \cite{Detweiler:1979wn} and binary black hole mergers \cite{LIGOScientific:2016aoc} has opened a new avenue for exploring gravity and the universe.
Recently, several collaborations within the pulsar timing array (PTA) community, including NANOGrav \cite{NANOGrav:2023gor,NANOGrav:2023hvm,NANOGrav:2020bcs}, EPTA \cite{EPTA:2023fyk}, PPTA \cite{Reardon:2023gzh} and CPTA \cite{Xu:2023wog}, have announced compelling evidence of a signal consistent with stochastic GW background at the frequency $f \simeq 1~ \textnormal{yr}^{-1}$.

Inflation, as the most popular paradigm of the early universe, predicts the existence of a primordial GW background \cite{Grishchuk:1974ny,Starobinsky:1979ty,Rubakov:1982df} across a wide frequency range ($10^{-18}-10^{10}$ Hz), yet to be confirmed by observations \cite{Guth:1980zm,Linde:1981mu,Albrecht:1982wi,Starobinsky:1980te}. It is interesting to ask whether the signals recently observed by PTA could originate from primordial GWs, see, e.g., \cite{Vagnozzi:2023lwo,Jiang:2023gfe,Zhu:2023lbf,Ye:2023tpz,Frosina:2023nxu,Ellis:2023oxs} (see also \cite{Huang:2023chx,Du:2023qvj,Wu:2023hsa,Xiao:2023dbb,Zhang:2023lzt,Ye:2023xyr,
King:2023ayw,Huang:2023zvs,Lozanov:2023rcd,Maji:2023fhv,Kawasaki:2023rfx,He:2023ado,
Choudhury:2023fwk,Choudhury:2023fjs,Jiang:2024dxj,DeAngelis:2024xtr}).
The standard slow-roll inflation predicts a nearly scale-invariant power spectrum of primordial GWs, rigorously constrained by Planck's data, which yields a tensor-to-scalar ratio $r_{0.002}<0.035$ within the observational window of the cosmic microwave background (CMB) \cite{BICEP:2021xfz}. However, PTA data indicates a highly blue-tilted GW power spectrum in the nanohertz frequency band, with a spectral index of $n_T=1.8\pm0.3$ \cite{NANOGrav:2023hvm,Vagnozzi:2023lwo}.

Attributing the observed signal in the PTA band to primordial GWs necessitates physics beyond standard slow-roll inflation, see, e.g., \cite{Jiang:2023gfe,Zhu:2023lbf,Ye:2023tpz,Borah:2023sbc,
Ben-Dayan:2023lwd,Oikonomou:2023qfz,Datta:2023xpr,Datta:2023vbs,Bhattacharya:2023ysp,Choudhury:2023hfm,Choudhury:2023kdb,Choudhury:2024one,Sharma:2024whg}, for recent studies. This would require the amplitude of the primordial GW power spectrum $P_T$ to reach a maximum of $P_T \simeq 10^{-3}$ at PTA scales (see, e.g., \cite{Ye:2023tpz}), where $P_T$ is related to the present GW energy density
spectrum $\Omega_{\rm{GW}}$ by $P_T\simeq 10^6 \Omega_{\rm{GW}}$. There have been numerous studies exploring the generation of a blue-tilted GW spectrum by introducing new physics beyond conventional slow-roll inflation, e.g., \cite{Piao:2004tq,Baldi:2005gk,Piao:2006jz,Kobayashi:2010cm,Kobayashi:2011nu,
Dudas:2010gi,Mourad:2017rrl,
Endlich:2012pz,Cai:2014uka,Gong:2014qga,Cannone:2014uqa,Wang:2014kqa,Kuroyanagi:2014nba,
Cai:2015yza,Cai:2016ldn,Wang:2016tbj,Fujita:2018ehq,Kuroyanagi:2020sfw,Akama:2020jko,
Akama:2023jsb,Giare:2020plo,Datta:2022tab,Giare:2022wxq,Oikonomou:2023bli,Choudhury:2023kam,
Oikonomou:2024aww}.
An intriguing direction is to introduce a violation of the null energy condition (NEC) during inflation.

Realizing a stable NEC violation is a challenge due to the presence of ghost or gradient instabilities in the primordial perturbations associated with the NEC violation \cite{Rubakov:2014jja,Libanov:2016kfc,Kobayashi:2016xpl,Easson:2011zy,Qiu:2015nha,Ijjas:2016tpn,Ijjas:2016vtq,Dobre:2017pnt,deCesare:2019pqj,deCesare:2021wmk}. It is first explicitly demonstrated in the framework of effective field theory that a fully stable\footnote{``Fully stable'' implies that both ghost and gradient instabilities are eliminated throughout the entire history of the universe.} NEC violation can be realized in ``beyond Horndeski'' theories \cite{Cai:2016thi,Creminelli:2016zwa,Cai:2017tku,Cai:2017dyi,Kolevatov:2017voe}, see also \cite{Cai:2017dxl,Cai:2017pga,Mironov:2018oec,Qiu:2018nle,Ye:2019frg,Ye:2019sth,Mironov:2019qjt,Akama:2019qeh,Mironov:2019mye,Ilyas:2020qja,Ilyas:2020zcb,Zhu:2021whu,Zhu:2021ggm,Mironov:2022ffa,Cai:2022ori,Panda:2024iqu}  for later developments. On this basis, it has been demonstrated that NEC violation during inflation can have significant observable effects, including: 1) an enhanced and blue-tilted power spectrum of primordial GWs with distinct features \cite{Cai:2020qpu,Cai:2022nqv}, which could be detected by PTA \cite{Ye:2023tpz}, Advanced LIGO and Advanced Virgo \cite{Chen:2024mwg}, and Taiji \cite{Chen:2024jca}, 2) the generation of primordial black holes and scalar-induced GWs \cite{Cai:2023uhc}, 3) an amplified parity-violation effect in primordial GWs \cite{Cai:2022lec,Zhu:2023lhv}.

In this paper, we investigate a new model where NEC violation occurs during slow-roll inflation, enabling the inflaton to climb over a high potential barrier before transitioning back to slow-roll inflation. Unlike the model in \cite{Cai:2020qpu}, in this model, the two stages of inflation before and after NEC violation have similar energy scales. We will investigate the observable effects of NEC violation in this model on the primordial GW background through analytical and numerical calculations.


\section{The model}\label{NECinf}

\begin{figure}[htbp]
    \includegraphics[scale=2,width=0.68\textwidth]{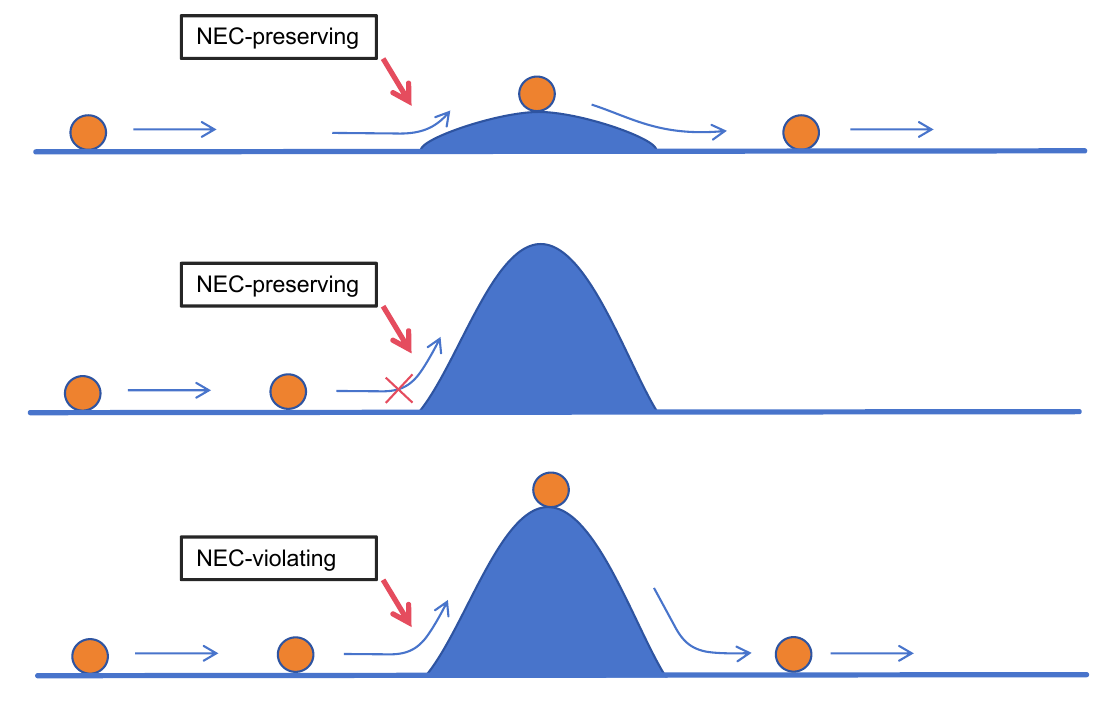}
\caption{In the first panel, inflaton can climb over the barrier which are not too high. But in the other two panels, inflaton requires the NEC violation to climb over the barrier and continue the slow-roll inflation.} \label{fig:ball}
\end{figure}

\begin{figure}[htbp]
    \includegraphics[scale=2,width=0.7\textwidth]{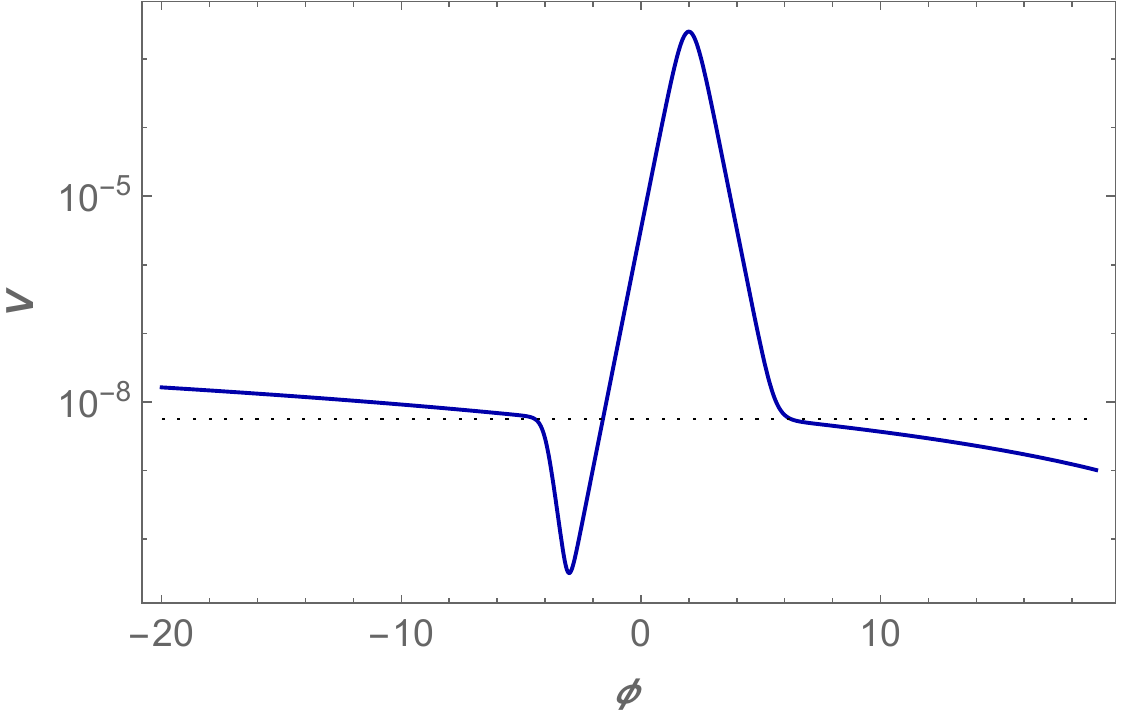}
\caption{A plot illustrating the potential $V(\phi)$ with parameters set as follows: $\phi_1=2$, $\phi_2=-4$, $\phi_4=20$, $\phi_5=29$, $q_2=6$, $q_4=4$, $\lambda=0.01$, $\sigma=23$, and $m=-4.5\times 10^{-6}$. The height of the potential barrier significantly exceeds that of both sides where slow-roll inflation occurs, i.e., $V_\mathrm{top}\gg V_\mathrm{inf1}$ and $V_\mathrm{inf2}$. As the scalar field rolls along the potential from left to right, the universe sequentially experiences the first stage of slow-roll inflation, NEC violation, and the second stage of slow-roll inflation.} \label{fig:figV-cropped}
\end{figure}

In our single scalar field model, the universe originates from a slow-roll inflation. As the inflaton rolls slowly down the potential, it encounters a high potential barrier (see Fig. \ref{fig:figV-cropped} for a sketch of the potential). Typically, a slowly rolling inflaton cannot climb over this high barrier. However, through the mechanism of NEC violation, the inflaton naturally violates the slow-roll conditions for a certain duration, allowing it to climb over the barrier and continue the slow-roll inflation, as illustrated by Fig. \ref{fig:ball}. Such an evolution will lead to interesting observable effects on the primordial perturbations.

Following \cite{Cai:2020qpu}, we use the action
\be\label{240417action}
S=\int d^4x\sqrt{-g}\Big[{M_p^2\over 2}R - M_p^2
    g_1(\phi)X/2 + g_2(\phi)X^2/4 -M_p^4 V(\phi) + L_{\delta g^{00} R^{(3)}} \Big] \,,
\ee
where $\phi$ is a dimensionless scalar field, $X=\nabla_{\mu}\phi\nabla^\mu\phi$, the EFT operator $L_{\delta g^{00} R^{(3)}}$ is adopted to thoroughly eliminate the ghost and gradient instabilities of the primordial scalar perturbations, see, e.g., \cite{Cai:2017dyi} for details. According to \cite{Cai:2016thi, Cai:2017dyi}, $L_{\delta g^{00} R^{(3)}}$ does not affect the background equations at all, nor does it contribute to the quadratic action of the tensor sector. It only modifies the squared sound speed of the scalar perturbations. For simplicity, we will focus on the primordial GWs in this paper.

The background equations can be obtained as
\ba \label{eqH}  3 H^2
M_p^2&=&\frac{M_p^2}{2} g_1\dot{\phi }^2 +\frac{3}{4} g_2 \dot{\phi
}^4 +M_p^4 V \,,
\\
\dot{H} M_p^2&=& -\frac{M_p^2}{2} g_1 \dot{\phi }^2 -\frac{1}{2} g_2
\dot{\phi }^4\,,  \label{dotH}
\\
0&=& \lf(g_1 +\frac{3 g_2 \dot{\phi }^2 }{M_p^2}\rt)\ddot{\phi}
+3 g_1 H\dot{\phi } +\frac{1}{2}g_{1,\phi }
\dot{\phi }^2 
+\frac{3 g_2 H\dot{\phi }^3}{M_p^2}  +\frac{3 g_{2,\phi } \dot{\phi }^4}{4M_p^2}
+M_p^2 V_{,\phi }\,, \label{eomphi} \ea where
``$_{,\phi}\equiv d/d\phi$'', a dot denotes the derivative with respect to the comic time $t$. In the background equations (\ref{eqH}) to (\ref{eomphi}), only two of them are independent.

\begin{figure}[htbp]
    \centering
\subfigure[~$g_1(\phi)$ ]
{\includegraphics[width=0.49\textwidth]{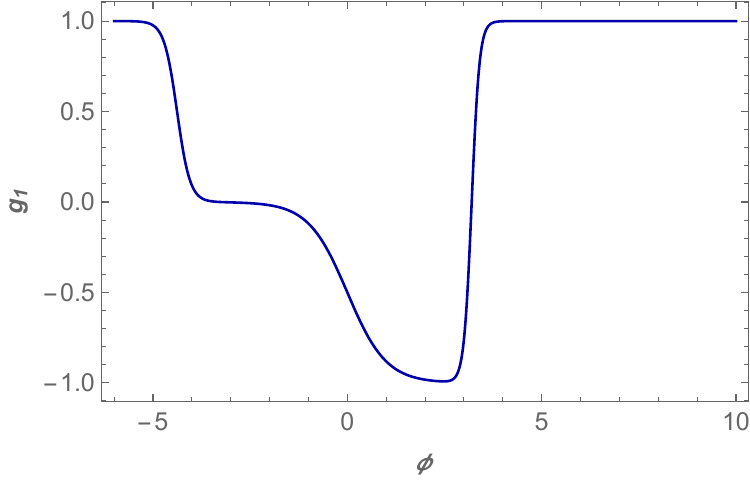}}
\subfigure[~$g_2(\phi)$ ]
{\includegraphics[width=0.48\textwidth]{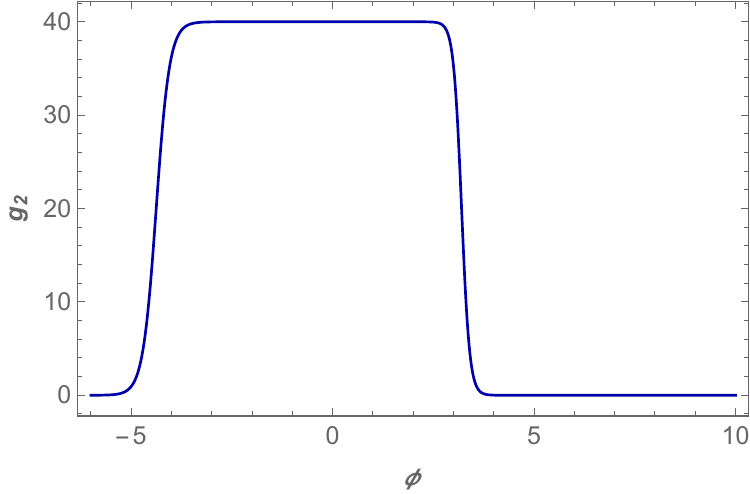}}
\caption{The functions $g_1(\phi)$ and $g_2(\phi)$, where we have set $q_1=10$, $q_2=6$, $q_3=10$, $\phi_0=3.2$, $\phi_3=-4.38$, $f_1=1$, $f_2=40$. A sufficiently large positive $g_2$ that satisfies $g_2{\dot \phi}^4 > M_{\rm P}^2 \dot{H}$ can prevent ghost instabilities.}\label{fig:g1g2}
\end{figure}

To achieve the aforementioned background evolution via NEC violation, we set these $\phi$-dependent functions $g_1$, $g_2$ and the potential $V$ in action (\ref{240417action}) as
\ba g_1(\phi)&=&
{1\over 1+e^{q_2(\phi-\phi_3)}} -{f_1 e^{2\phi} \over 1+f_1
e^{2\phi} }+{2\over 1+e^{-q_1(\phi-\phi_0)}}\,,\label{g1}
\\
g_2(\phi)&=& {f_2\over 1+e^{-q_2(\phi-\phi_3)}} {1\over
    1+e^{q_3(\phi-\phi_0)}}\,,
\label{g3}
\ea
\begin{equation}
\begin{split}
V(\phi)=&{1\over2}m^2(\phi-\phi_4)^2 {1\over
    1+e^{q_2(\phi-\phi_2)}}+
\lambda\left[1-\frac{(\phi-\phi_1)^{2}}{\sigma^{2}}\right]^{2}{e^{-q_4(\phi-\phi_1)}\over
    {(1+e^{-q_4(\phi-\phi_1)})}^2}\\
    &+{1\over2}m^2(\phi-\phi_5)^2 {1\over
    1+e^{-q_4(\phi-\phi_1)}}\,, \label{V}
\end{split}
\end{equation}
where $\lambda$, $m$, $f_{1,2}$, $q_{1,2,3,4}$ and $\phi_{0,1,2,3,4,5}$ are dimensionless constants. We plot $g_1(\phi)$, $g_2(\phi)$ and $V(\phi)$ for appropriate values of these parameters in Figs. \ref{fig:g1g2} and \ref{fig:figV-cropped}.
As the scalar field rolls along the potential from left to right, the universe sequentially experiences the first stage of slow-roll inflation, NEC violation, and the second stage of slow-roll inflation. These two slow-roll stages are labeled as ``inf1'' and ``inf2'' respectively.

The parameters introduced above should satisfy $\phi_3<\phi_2<0<\phi_1<\phi_0$. As can be seen in Figs. \ref{fig:g1g2} and \ref{fig:figV-cropped}, when $\phi$ is much smaller than $\phi_3$, $g_1=1$ and $g_2=0$, rendering the potential responsible for the first stage of slow-roll inflation. In this regime, the first term of the potential predominates. Conversely, as $\phi$ greatly exceeds $\phi_0$, $g_1=1$ and $g_2=0$, and the potential is responsible for the second stage of slow-roll inflation, with the third term in the potential becoming dominant. The exponential term in each expression serves to modulate the dominance of various terms across different regions of the potential.

From Fig. \ref{fig:figV-cropped}, it can be observed that the height of the potential barrier significantly exceeds that of both sides where slow-roll inflation occurs. For a slowly rolling canonical scalar field, such a high barrier is insurmountable. However, with the action described in Eq. (\ref{240417action}) and via the NEC violation mechanism, the scalar field can naturally disrupt the slow-roll conditions and rapidly climb over the barrier.

The drop of the potential before the hill is significant for the $\phi$ field to climb to the top of the potential. In our model, the equation of motion for $\phi$ is given by Eq. (\ref{eomphi}). The coefficient in front of $\ddot{\phi}$ is proportional to $Q_s H^2/\dot{\phi}^2$, which must remain positive throughout to avoid ghost instability. To allow the $\phi$ field to climb to the top of the potential without requiring a non-zero initial kinetic energy, we set the model parameters so that $g_1 < 0$ around the NEC-violating phase. This results in the term $3g_1 H\dot{\phi}$ acting as an anti-friction force, thereby accelerating the $\phi$ field. Therefore, both the magnitude of $\dot{\phi}$ at the beginning of the climb and the coefficient in front of $\ddot{\phi}$ will influence the $\phi$ field's ability to climb over the hill.
This highlights the importance of the drop of the potential before the hill. Otherwise, if $\dot{\phi}$ is too small when the field starts climbing, we might need to finely tune the model parameters to ensure that the $\phi$ field reaches the top of the potential.

\begin{figure}[htbp]
    \centering
\subfigure[ ]
{\includegraphics[width=0.47\textwidth]{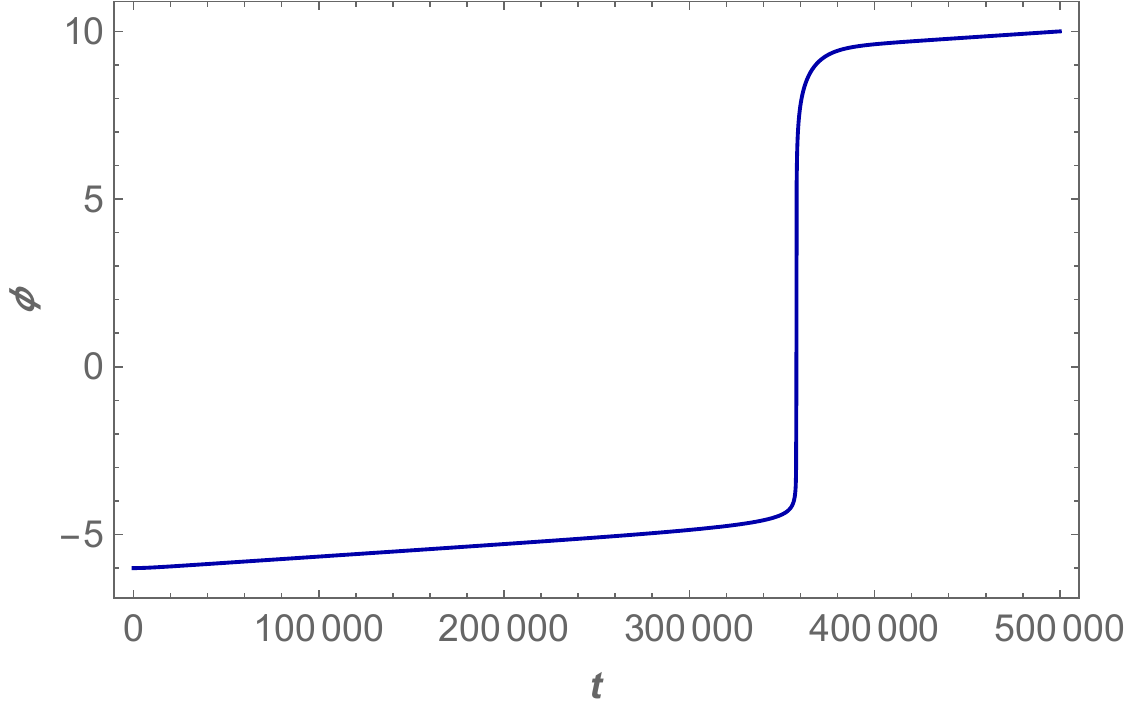}\label{fig:figphi-num}}
    \quad
\subfigure[ ]
{\includegraphics[width=0.47\textwidth]{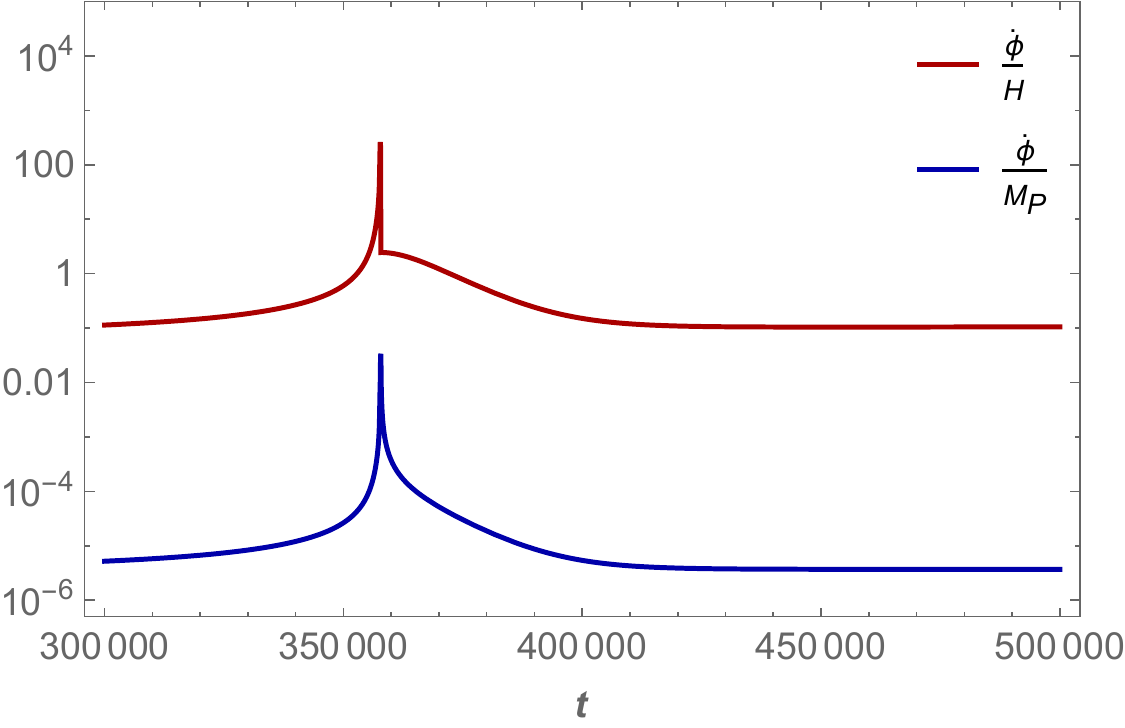}\label{fig:dotfigphi-num}}
\caption{The evolutions of $\phi(t)$ (left panel) and $\dot{\phi}$ (right panel), where $\phi(t_{\rm ini})=-6$, $\dot{\phi}(t_{\rm ini})=0$, $t_{\rm ini}=0$, $\phi_0=3.2$, $\phi_1=2$, $\phi_2=-4$, $\phi_3=-4.38$, $q_1=10$, $q_2=6$, $q_3=10$, $q_4=4$,
$f_1=1$, $f_2=40$, $\lambda=0.01$, $\sigma=23$ and $m=-4.5\times10^{-6}$.} \label{fig:phidotphi}
\end{figure}

\begin{figure}[htbp]
\centering
    \subfigure[ ]
    {\includegraphics[width=.46\textwidth]{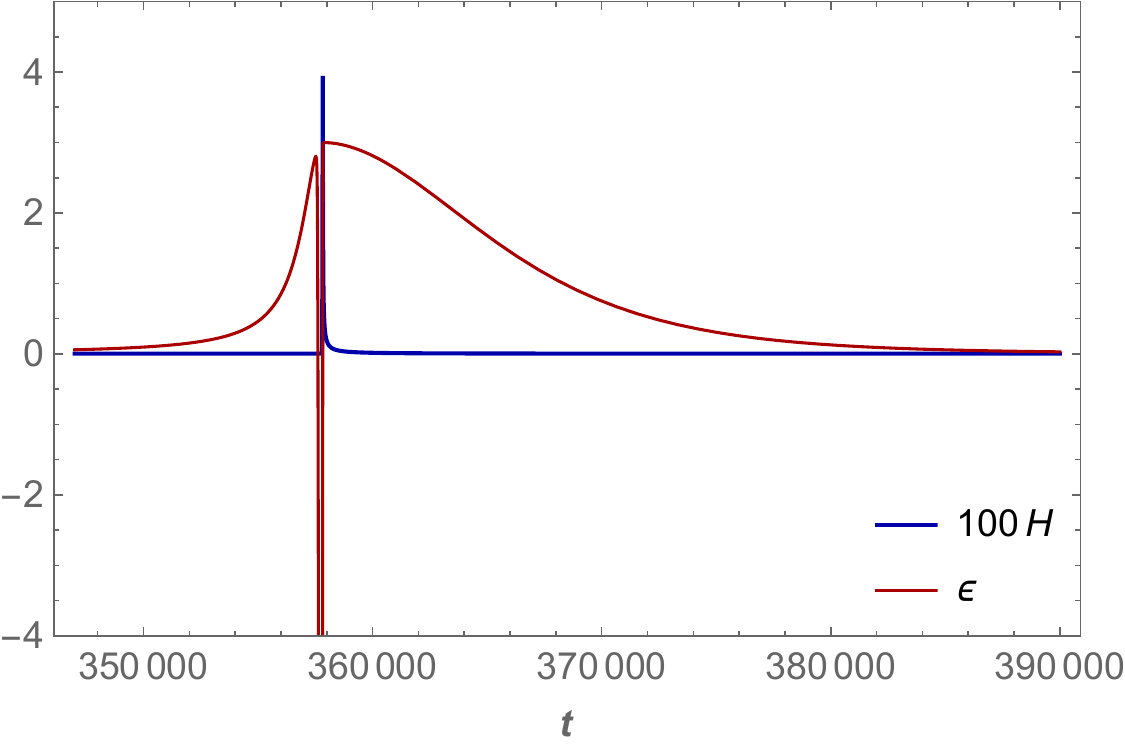} \label{fig:H}}
    \subfigure[ ]
    {\includegraphics[width=.46\textwidth]{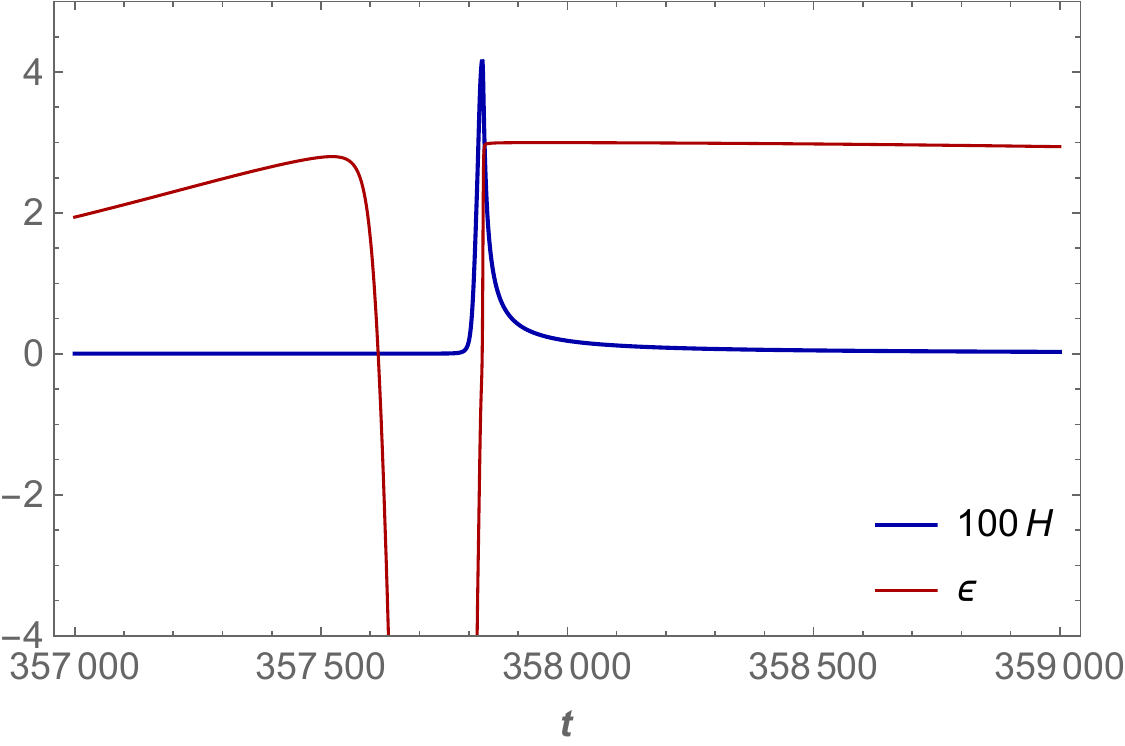} \label{fig:Hbig}}
    \caption{The evolutions of a hundredfold of $H$ (blue curve) and $\epsilon=-\dot{H}/H^2$ (red curve). We have $0<\epsilon\ll1$ during the first and second stages of slow-roll inflation, and $\epsilon\ll-1$ during the NEC-violating stage. Note that when $\dot{\phi}/H$ reaches its peak, $H$ is still far from its maximum.} \label{fig:H1}
\end{figure}

We numerically solve the background evolutions and display the evolutions of $\phi$, $\dot{\phi}$, $H$ and $\epsilon\equiv-\dot{H}/H^2$ in Figs. \ref{fig:phidotphi} and \ref{fig:H1}. From these figures, we can see that the universe sequentially experiences the first stage of slow-roll inflation, NEC violation, and the second stage of slow-roll inflation. This background evolution is similar to that in the scenario described in \cite{Cai:2020qpu}. The main difference from the model in \cite{Cai:2020qpu} lies in the fact that in this model, the energy scales of the first and second stages of slow-roll inflation are closer to each other.

\section{The power spectrum of Primordial GWs}

In this section, we calculate the power spectrum of primordial GWs for our model. The quadratic actions of tensor perturbation mode $\gamma_{ij}$ can be given as
\be S^{(2)}_{\gamma}={M_p^2\over8}\int d^4xa^3 \lf[
\dot{\gamma}_{ij}^2 -{(\partial_k\gamma_{ij})^2\over
    a^2}\rt]\,.\label{tensor-action}
\ee
In the momentum space,
\be
\gamma_{ij}(\tau,\mathbf{x})=\int \frac{d^3k}{(2\pi)^{3}
}e^{-i\mathbf{k}\cdot \mathbf{x}} \sum_{\lambda=+,\times}
\hat{\gamma}_{\lambda}(\tau,\mathbf{k})
\epsilon^{(\lambda)}_{ij}(\mathbf{k}), \ee where
$\hat{\gamma}_{\lambda}(\tau,\mathbf{k})=
\gamma_{\lambda}(\tau,k)a_{\lambda}(\mathbf{k})
+\gamma_{\lambda}^*(\tau,-k)a_{\lambda}^{\dag}(-\mathbf{k})$,
$\epsilon_{ij}^{(\lambda)}(\mathbf{k})$ satisfies
$k_{j}\epsilon_{ij}^{(\lambda)}(\mathbf{k})=0$,
$\epsilon_{ii}^{(\lambda)}(\mathbf{k})=0$,
$\epsilon_{ij}^{(\lambda)}(\mathbf{k})
\epsilon_{ij}^{*(\lambda^{\prime}) }(\mathbf{k})=\delta_{\lambda
    \lambda^{\prime} }$ and $\epsilon_{ij}^{*(\lambda)
}(\mathbf{k})=\epsilon_{ij}^{(\lambda) }(-\mathbf{k})$;
$a_{\lambda}(\mathbf{k})$ and
$a^{\dag}_{\lambda}(\mathbf{k}^{\prime})$ satisfy $[
a_{\lambda}(\mathbf{k}),a_{\lambda^{\prime}}^{\dag}(\mathbf{k}^{\prime})
]=\delta_{\lambda\lambda^{\prime}}\delta^{(3)}(\mathbf{k}-\mathbf{k}^{\prime})$.

Using the Eq. (\ref{tensor-action}), we can obtain the equation of motion for ${\gamma}_{\lambda}(\tau,\mathbf{k})$ as
\be u_k^{\prime\prime}+\left(k^{2}-\frac{a^{\prime \prime}}{a}\right)
u_k=0\,,\label{eq:eomu}
\ee
where $u_k\equiv \gamma_{\lambda}(\tau,k)aM_p/2$ and a prime denotes $d/d\tau$.
The power spectrum of primordial GWs is defined as
\be
P_T={k^3\over 2\pi^2}\sum_{\lambda=+,\times}|\gamma_\lambda|^2=\frac{4k^3}{\pi^2
M_p^2}\cdot\frac{\lf|u_{k}
    \rt|^2}{ a^2}\,,\label{eq:PT01}
\ee
which should be evaluated at the end of the second stage of slow-roll inflation.

\begin{figure}[htbp]
\includegraphics[width=0.7\textwidth]{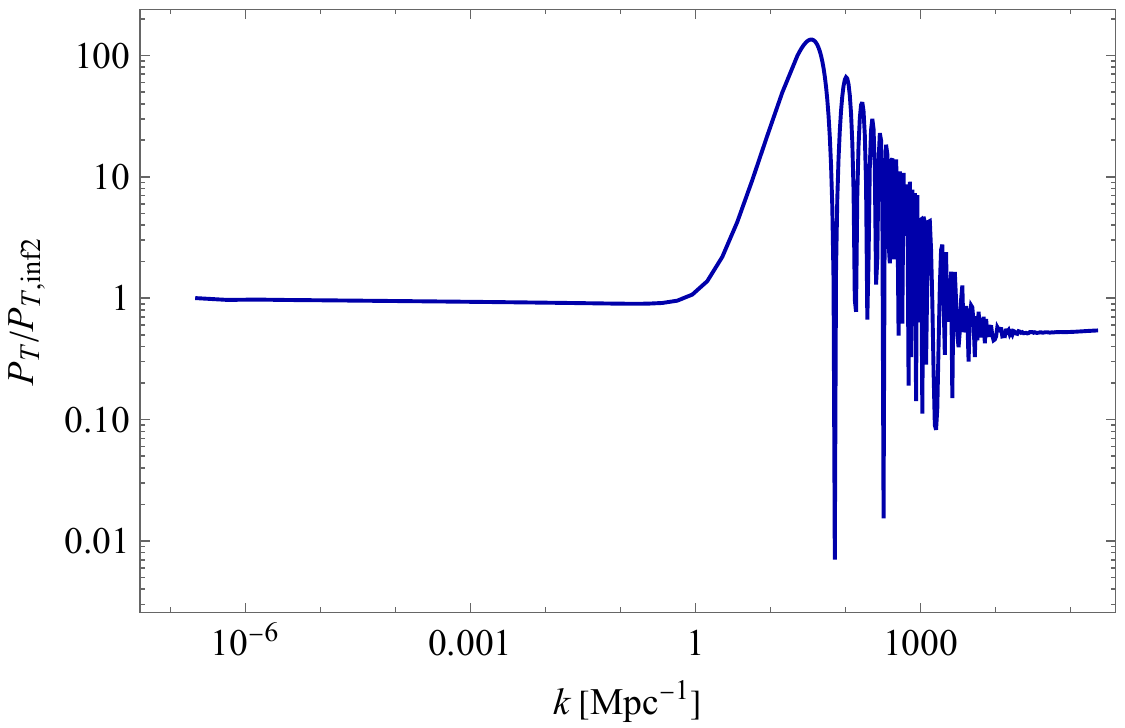}
\caption{The power spectrum $P_T$ of primordial GWs for our model obtained by numerically solving Eq. (\ref{eq:eomu}). We have set $q_1=10$, $q_2=6$, $q_3=10$, $q_4=4$, $\phi_0=3.2$, $\phi_1=2$, $\phi_2=-4$, $\phi_3=-4.38$, $\phi_4=20$, $\phi_5=29$, $f_1=1$, $f_2=40$, $\lambda=0.01$, $\sigma=23$, $m=-4.5\times
10^{-6}$, $\phi(t_{\rm ini})=-6$, $\dot{\phi}(t_{\rm ini})=0$ and $t_{\rm ini}=0$.}\label{fig:PT001}
\end{figure}

We numerically solve Eq. (\ref{eq:eomu}) and obtain the power spectrum of primordial GWs for our model. The results are displayed in Fig. \ref{fig:PT001}. We observe a significant enhancement in the power spectrum amplitude of perturbation modes that exit the horizon around the NEC violation epoch in this model. This is because perturbation modes that are outside the horizon during the NEC violation period experience rapid growth, surpassing the amplitude of the constant mode.
Therefore, the power spectrum at intermediate scales exhibits a blue tilt. Similar to the model in Ref. \cite{Cai:2020qpu}, the model in this paper also has the potential to explain observations at the PTA scale. However, unlike the model in Ref. \cite{Cai:2020qpu}, the power spectrum amplitudes of perturbation modes that exit the horizon during the second stage of slow-roll inflation are not significantly different from those that exit during the first stage.

In the following, we attempt to provide an analytical derivation of the power spectrum. Analogous to the approach in Ref. \cite{Cai:2020qpu}, we divide the background evolution in this model into four stages, with the scale factor of the $j$-th stage approximately parameterized as
\be
a_{j}(\tau)\sim
\left({\tau}_{R,j}-\tau\right)^{\frac{1}{\epsilon_{j}-1}}, \label{eq:aparaj}
\ee
where $\epsilon_j\equiv -\dot{H_j}/H_j^2={3\over 2}(1+w_j)$, ${\tau}_{R,j} \equiv\tau_{j}-(\epsilon_{j}-1)^{-1} \mathcal{H}^{-1}(\tau_j)$, $j=1$, 2, 3, 4. For simplicity, we assume $\epsilon_j$ (i.e., the equation of state parameter $w_j$) as a constant. We also require continuity of $a$ and $a'$ at each transition moment.

With Eq. (\ref{eq:aparaj}), we have
\be
\frac{a_j^{\prime
        \prime}}{a_j}={\nu_j^2-{1/4} \over
    {(\tau-\tau_{R,j})^2}}\,,\label{zppbz02} \ee where
$\nu_j={3\over2}\lf|{1-w_j\over 1+3 w_j}\rt|$. Consequently, the solutions to Eq. (\ref{eq:eomu}) for each phase can be given as
\ba u_{k,1}(\tau)&=&{\sqrt{\pi
		(\tau_{R,1}-\tau)}\over 2}\lf\{\alpha_{1}H_{\nu_{1}}^{(1)}[k(\tau_{R,1}-\tau)]+\beta_{1}H_{\nu_1}^{(2)}[k(\tau_{R,1}-\tau)]
\rt\} ,\,\, (\tau<\tau_{1})\,,
\\
u_{k,{2}}(\tau)&=&{\sqrt{\pi (\tau_{R,{2}}-\tau)}\over
	2}\Big\{\alpha_{2}H_{\nu_{2}}^{(1)}[k(\tau_{R,2}-\tau)]
+\beta_{2}H_{\nu_{2}}^{(2)}[k(\tau_{R,2}-\tau)] \Big\}
, (\tau_1\leq\tau\leq\tau_{2})\,,
\nn
\\
u_{k,{3}}(\tau)&=&{\sqrt{\pi (\tau_{R,{3}}-\tau)}\over
	2}\Big\{\alpha_{3}H_{\nu_{3}}^{(1)}[k(\tau_{R,3}-\tau)]
+\beta_{3}H_{\nu_{3}}^{(2)}[k(\tau_{R,3}-\tau)] \Big\}
, (\tau_2<\tau \leq \tau_{3})\,,\nn
\\
u_{k,{4}}(\tau)&=&{\sqrt{\pi (\tau_{R,{4}}-\tau)}\over
	2}\Big\{\alpha_{4}H_{\nu_{4}}^{(1)}[k(\tau_{R,4}-\tau)]
+\beta_{4}H_{\nu_{4}}^{(2)}[k(\tau_{R,4}-\tau)] \Big\}
, (\tau_3<\tau \leq \tau_{4})\,,\nn
\label{solution}
\ea
where $\alpha_{j}$ and $\beta_{j}$ are $k$-dependent coefficients, $j=1$, 2, 3, 4.
We set the initial state as the Bunch-Davis vacuum, i.e., $u_k= e^{-i k\tau}/\sqrt{2 k}$, which indicates that $|\alpha_1|=1$ and $|\beta_1|=0$.

\begin{figure}[htbp]
    \centering
\subfigure[ ]
{\includegraphics[width=0.48\textwidth]{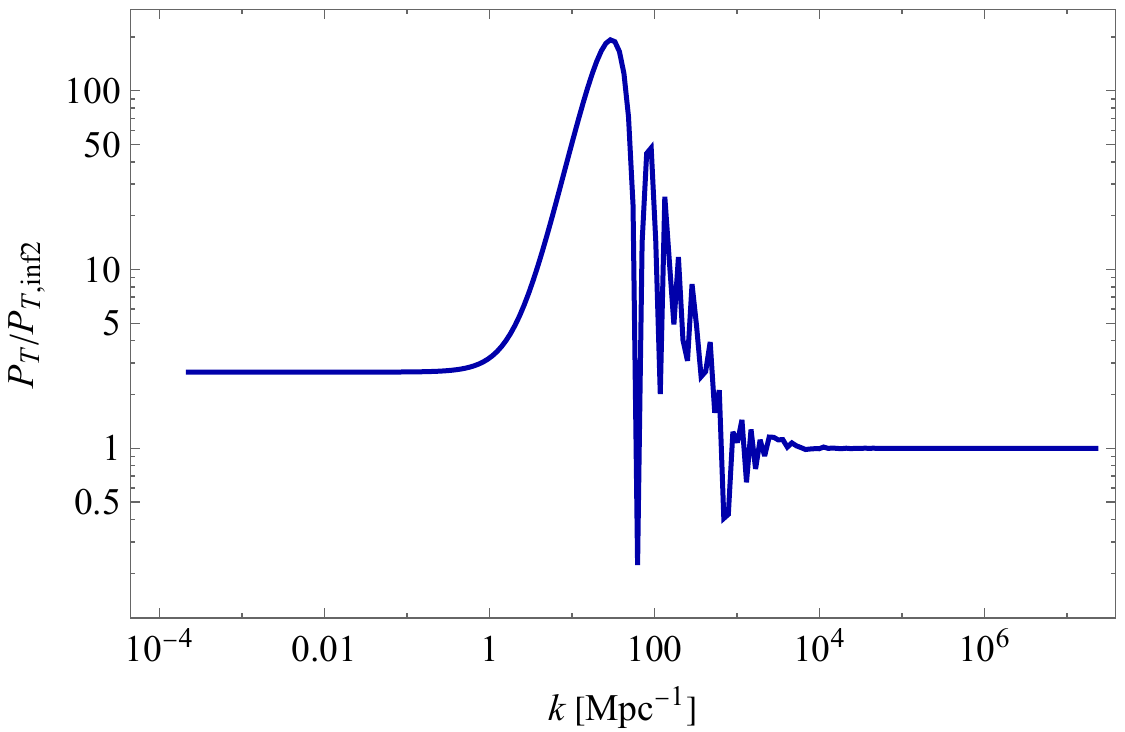}\label{fig:pt-num-omiga3-1}}
    \quad
\subfigure[ ]
{\includegraphics[width=0.47\textwidth]{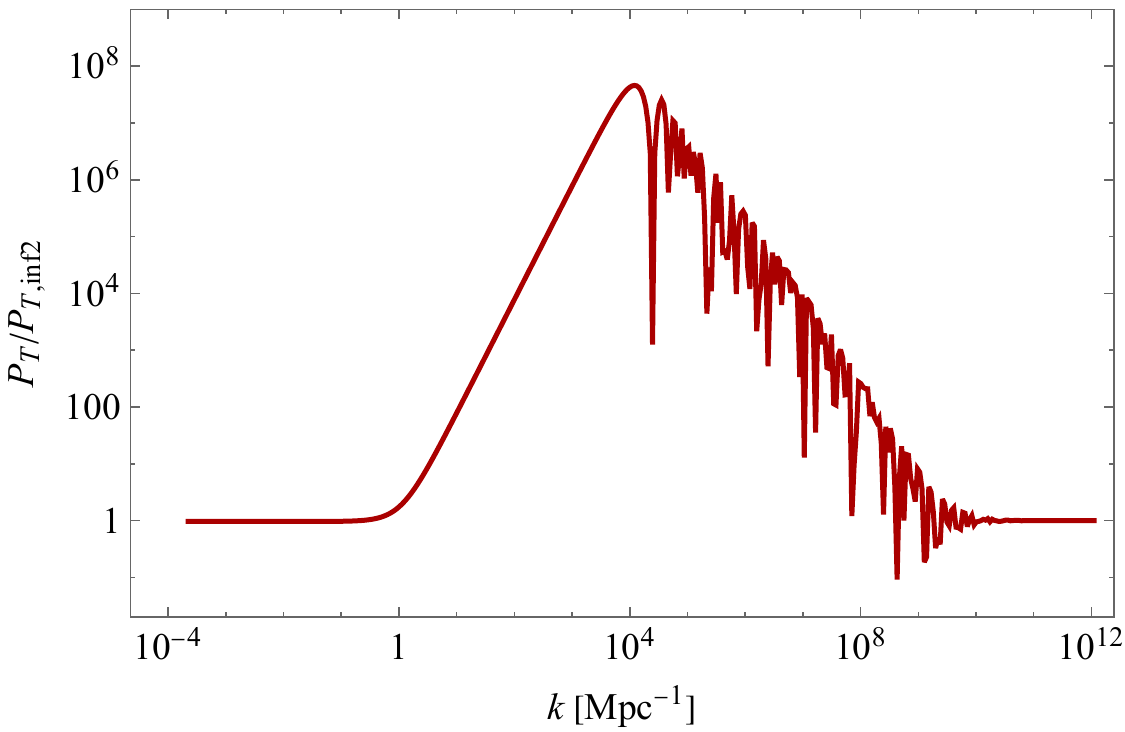}\label{fig:pt-num-3omiga3-2}}
\caption{The power spectrum of the primordial GWs given by Eq. (\ref{eq:PT}).
For the left panel, we set $w_1=-1$, $w_2=-10$, $w_3=1$, and $w_4=-1$, providing a close approximation to the result shown in Fig. \ref{fig:PT001}. In the right panel, we choose $w_1=-1$, $w_2=-8$, $w_3=2/3$, and $w_4=-1$, resembling the scenario discussed in Ref. \cite{Cai:2020qpu}. }\label{fig:PT-002}
\end{figure}

Using the matching conditions $u_{k,j}(\tau_{j})=u_{k,j+1}(\tau_{j})$ and $u_{k,j}' \big|{\tau=\tau{j}}= u_{k,j+1}' \big|{\tau=\tau{j}}$, we can obtain all of the $\alpha_{j}$ and $\beta_{j}$ for $j= 2$, 3, 4.
We have
\ba \left(
\begin{array}{ccc} \alpha_{j+1}\\ \beta_{j+1}
\end{array}\right)
&=& {\cal M}^{(j)}
\left(\begin{array}{ccc} \alpha_{j}\\
    \beta_{j}\end{array}\right)\,,\quad {\rm where} \quad
{\cal M}^{(j)} = \left(\begin{array}{ccc}
    {\cal M}^{(j)}_{11}&{\cal
        M}^{(j)}_{12}\\
    {\cal M}^{(j)}_{21}&{\cal M}^{(j)}_{22}\end{array}\right)\,,
 \label{Mmetric}
\ea
The details of the matrix elements of the transfer matrix can be found in \cite{Cai:2015nya,Cai:2019hge}.
At the end of the second slow-roll inflation, the power spectrum can be given as
\be P_T=\frac{4k^3}{\pi^2
M_p^2}\cdot\frac{\lf|u_{k,4}
    \rt|^2}{ a^2}\Big|_{\tau\rightarrow\tau_4}= P_{T,inf2} \lf|\alpha_4 - \beta_4
\rt|^2\,,\label{eq:PT}
\ee
where $P_{T,inf2}={2H_{inf2}^2 \over \pi^2 M_p^2}$.
A detailed expression for $\alpha_4$ and $\beta_4$ can be found in Appendix \ref{Sec:app1}. We plot $P_T/P_{T,\text{inf2}}=\left|\alpha_4 - \beta_4 \right|^2$ in Fig. \ref{fig:PT-002} for different choices of $w_i$. We can see from Figs. \ref{fig:PT001} and \ref{fig:pt-num-omiga3-1} that the approximation works well for our model with appropriate choices of $w_i$.
In our parameterization, the height difference between the two platforms in the potential can be adjusted by changing $\epsilon_3=-\dot{H}/H^2={3\over 2}(1+w_j)$.

\section{The energy density spectrum of GWs}

\begin{figure}[htbp]
\includegraphics[width=0.7\textwidth]{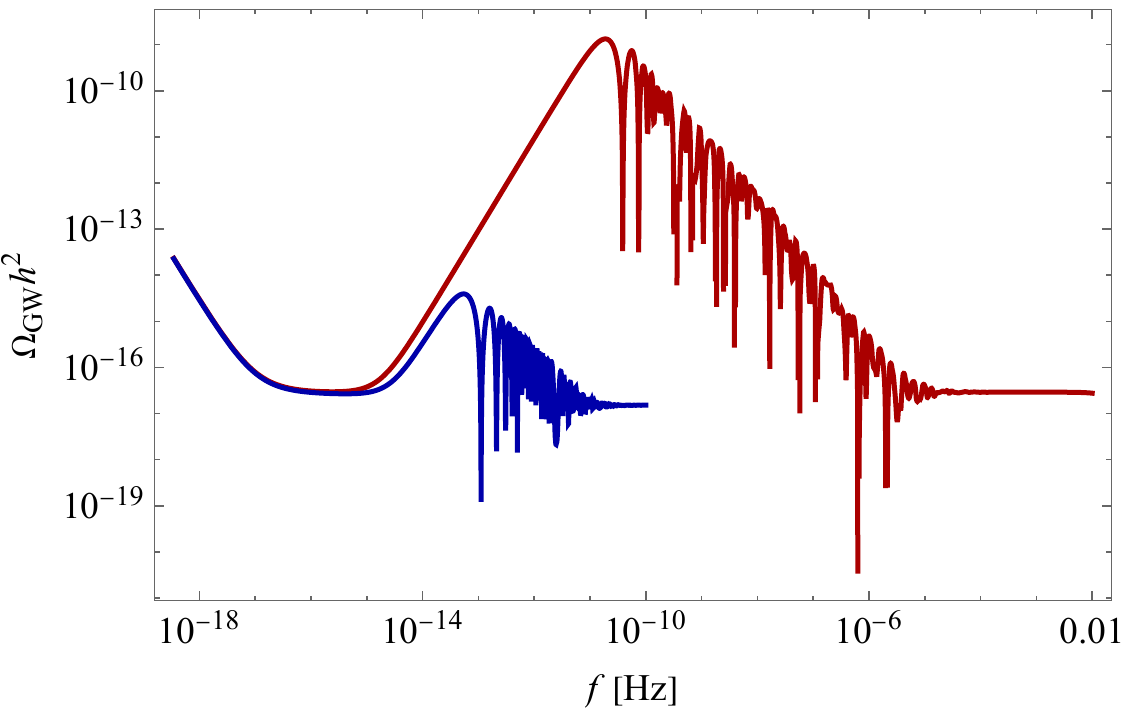}
\caption{ The energy density spectrum of GWs for our model. The blue curve is obtained by using the numerical results obtained by solving Eq. (\ref{eq:eomu}), while the red curve is based on the power spectrum shown in Fig. \ref{fig:pt-num-3omiga3-2}. We have set $P_{T,inf2}=1.82\times10^{-10}$.}\label{fig:numenergy}
\end{figure}

The energy density spectrum of GWs \cite{Turner:1993vb,Boyle:2005se} (see also \cite{Zhao:2006mm,Kuroyanagi:2014nba}) is
 \be
\Omega_{\rm{GW}}(\tau_{0})=\frac{k^{2}}{12
a_0^2H^2_0}P_{T}(k)\lf[\frac{3
\Omega_{{m}}j_1(k\tau_0)}{k\tau_{0}}\sqrt{1.0+1.36\frac{k}{k_{\text{eq}}}
+2.50\left( \frac{k}{k_{\text{eq}}}\right) ^{2}}\rt]^2,
\label{GW0}
\ee
where $H_0=67.8\, {\rm km/s/Mpc}$, $\tau_{0}=1.41\times10^{4}$ Mpc, $a_0=1$, $k_{\text{eq}}=0.073\,\Omega_{\text{m}} h^{2}$
Mpc$^{-1}$ is the wavenumber corresponding to the mode that entered the horizon at the equality of matter and radiation, $\Omega_{\rm m}=\rho_m/\rho_c$ is the density fraction of matter today, $\rho_{{c}}=3H^{2}_0/\big(8\pi
G\big)$ is the critical energy density, and $j_1(x)$ is the spherical Bessel function of the first kind.
We plot the energy density spectrum of GWs for our model in Fig. \ref{fig:numenergy}, using the numerical results obtained by solving Eq. (\ref{eq:eomu}).
Namely, Fig. \ref{fig:numenergy} correspond to the power spectrum displayed in Fig. \ref{fig:PT001}.

For observational purposes, the distinctive features may include the blue-tilted nature in the middle-frequency range and the decreasing amplitude of these oscillations at higher frequencies. Recently, observational data reported by NANOGrav \cite{NANOGrav:2023gor,NANOGrav:2023hvm,NANOGrav:2020bcs}, EPTA \cite{EPTA:2023fyk}, PPTA \cite{Reardon:2023gzh} and CPTA \cite{Xu:2023wog} is consistent with a stochastic GW background with an amplitude of $\Omega_{\rm{GW}} \simeq 10^{-9}$ and a spectral index of $n_T = 1.8 \pm 0.3$ at the reference frequency $f=1$ $\text{yr}^{-1}$. If the observational data are interpreted as primordial GW background, it corresponds to a strongly blue-tilted spectrum with an amplitude of $P_T \simeq 10^6 \Omega_{\rm{GW}}\simeq 10^{-3}$ at a frequency of $f = 1$ $\text{yr}^{-1}$. We have plotted the energy density spectrum in red in Fig. \ref{fig:numenergy} using the analytical formula given in Eq. (\ref{eq:PT}) for a specific set of parameters. Fig. \ref{fig:numenergy} suggests that future PTA observations may have the potential to detect or constrain our model through the specific features of the power spectrum, which requires further investigation.


\section{Conclusion}

In this paper, we propose a new single-field model where the universe sequentially undergoes a first stage of slow-roll inflation, NEC violation, and a second stage of slow-roll inflation. Within the potential responsible for slow-roll inflation, there exists a high barrier, over which the rolling inflaton climb via the NEC violation mechanism\footnote{See also \cite{Caravano:2024tlp} for other mechanism.} before entering the second inflationary stage. Following the NEC violation, there is a process of energy scale reduction, resulting in the energy scales of the two slow-roll inflation stages being relatively close, contrasting with the model in \cite{Cai:2020qpu}.

We calculate the power spectrum of primordial GWs using both numerical and analytical methods, comparing the results obtained from these two approaches. We find that the approximation used in the analytical derivation closely matches the results from numerical calculations.
The power spectrum is nearly scale-invariant on both large and small scales. However, in the middle-frequency range, it shows a blue-tilted nature, while at higher frequencies, there is a reduction in oscillation amplitudes. These unique characteristics could potentially differentiate our model from others.

\begin{figure}[htbp]
    \includegraphics[scale=2,width=0.55\textwidth]{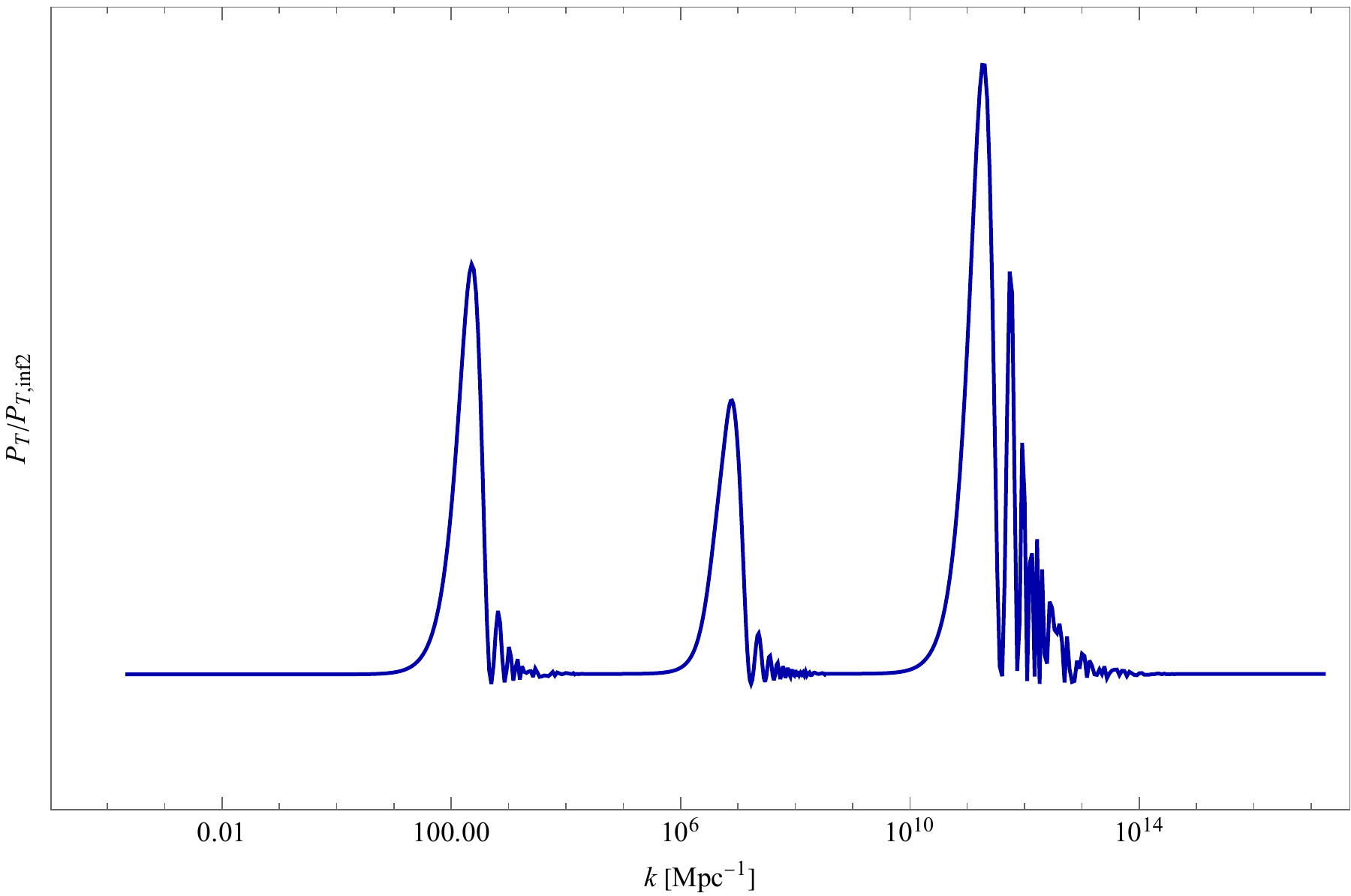}
\caption{A sketch of the GW power spectrum for the scenario where $\phi$ climbs over the potential barriers multiple times.} \label{fig:duojieduan}
\end{figure}

In the early universe, NEC violation may have occurred multiple times. We present a sketch of the GW power spectrum in Fig. \ref{fig:duojieduan} for the scenario where $\phi$ climbs over the potential barriers multiple times. The accumulation of future observational data from various frequency bands, including CMB, PTA, Advanced LIGO, Advanced Virgo, LISA, Taiji and Tianqin, will enable more tests or constraints on NEC violation in the very early universe.

\acknowledgments

Y. C. is supported in part by the Natural Science Foundation of Henan Province and Zhengzhou University (Grant Nos. 242300420231, JC23149007, 35220136), the China Postdoctoral Science Foundation (Grant No. 2021M692942). Y.-S. P. is supported by NSFC, No.12075246, National Key Research and Development Program of China, No. 2021YFC2203004, and the Fundamental Research Funds for the Central Universities.

\appendix

\section{Detailed analytic expression of $\alpha$ and $\beta$}\label{Sec:app1}

The scale factor can be parameterized as a piecewise function of $\tau$, i.e.,
\begin{equation}
\label{eq:atau}
    a_j(\tau) = a_j (\tau_j) \left( \frac{\tau - \tau_{R,j}}{\tau_j - \tau_{R,j}} \right)^{\frac{1}{\epsilon_j - 1}} ~,\quad \tau<\tau_j \,.
\end{equation}
The continuity of $a$ gives
\begin{equation}
    a_1(\tau_1) = a_2(\tau_2) \left( \frac{\tau_1 - \tau_{R,2}}{\tau_2 - \tau_{R,2}} \right)^{\frac{1}{\epsilon_2 - 1}},\end{equation}
\begin{equation}
    a_2(\tau_2) = a_3(\tau_3) \left( \frac{\tau_2 - \tau_{R,3}}{\tau_3 - \tau_{R,3}} \right)^{\frac{1}{\epsilon_3 - 1} },~\end{equation}
\begin{equation}
    a_3(\tau_3) = a_4(\tau_4) \left( \frac{\tau_3 - \tau_{R,4}}{\tau_4 - \tau_{R,4}} \right)^{\frac{1}{\epsilon_4 - 1} }.\end{equation}
Considering the continuity of $a'$ or $\mathcal{H}$, we can define the following quantities:
\begin{equation}
    \bar{\mathcal{H}}_1 \equiv \mathcal{H}(\tau_1) ~,~ \bar{\mathcal{H}}_2 \equiv \mathcal{H}(\tau_2) ~,~
    \bar{\mathcal{H}}_3 \equiv \mathcal{H}(\tau_3) ~.
\end{equation}
Using \eqref{eq:atau}, we can obtain the integration constants $\tau_{R,j}$ as
\begin{equation}
\label{eq:Hjunction}
    \tau_{R,1} = \tau_1 + \bar{\mathcal{H}}_1^{-1} ~,~ \tau_{R,2} = \tau_1 + ({v_2-\frac{1}{2}}){\bar{\mathcal{H}}_1^{-1}}, \tau_{R,2} = \tau_2 + ({v_2-\frac{1}{2}}){\bar{\mathcal{H}}_2^{-1}},\end{equation}
\begin{equation}
    \tau_{R,3} = \tau_2 + ({-v_3-\frac{1}{2}}){\bar{\mathcal{H}}_2^{-1}},\tau_{R,3} = \tau_3 + ({-v_3-\frac{1}{2}}){\bar{\mathcal{H}}_3^{-1}},~ \tau_{R,4} = \tau_3 + \bar{\mathcal{H}}_3^{-1} ,
\end{equation}where
$\nu_j={3\over2}\lf|{1-w_j\over 1+3 w_j}\rt|$, and $\epsilon_j={3\over 2}(1+w_j)\simeq {\rm const.}$. Note that for $-{1\over3}<w_j<1$ and $w_j<-{1\over3}$, we have $\nu_j=-{1\over2}-{1\over1-\epsilon_j}$ and $\nu_j={1\over2}+{1\over1-\epsilon_j}$, respectively.

Using the matching method, we obtain the following recursive relations:
\begin{align}
    ~~~~~~\alpha_2 &= \frac{i \pi }{8 x_1} \sqrt{\frac{x_1}{x_2}} \left( H_{\frac{3}{2}}^{(1)}(x_1) \left( 2 x_1 x_2 H_{\nu_2-1}^{(2)}(x_2) + \left(-2 \nu_2 x_1+x_1+2 x_2\right) H_{\nu_2}^{(2)}(x_2) \right) \right. \notag \\
    & \quad - 2 x_1 x_2 H_{\frac{1}{2}}^{(1)}(x_1) H_{\nu_2}^{(2)}(x_2) \Big),
\end{align}
\begin{align}
    ~~~~~~~~~\beta_2 &= -\frac{i \pi }{8 x_1} \sqrt{\frac{x_1}{x_2}} \left( H_{\frac{3}{2}}^{(1)}(x_1) \left( 2 x_1 x_2 H_{\nu_2-1}^{(1)}(x_2) + \left(-2 \nu_2 x_1+x_1+2 x_2\right) H_{\nu_2}^{(1)}(x_2) \right) \right. \notag \\
    & \quad - 2 x_1 x_2 H_{\frac{1}{2}}^{(1)}(x_1) H_{\nu_2}^{(1)}(x_2) \Big),
\end{align}
\begin{align}
    \alpha_3 &= -\frac{\pi}{8 y_1} \sqrt{-\frac{y_1}{y_2}} \Big( y_2 H_{\nu_3}^{(2)}(y_2) \Big( 2 \alpha_2 y_1 H_{\nu_2-1}^{(1)}(y_1) + (\alpha_2-2 \alpha_2 \nu_2) H_{\nu_2}^{(1)}(y_1) \notag \\
    & \quad + 2 \beta_2 y_1 H_{\nu_2-1}^{(2)}(y_1) + (\beta_2-2 \beta_2 \nu_2) H_{\nu_2}^{(2)}(y_1) \Big) \notag \\
    & \quad - y_1 \Big( 2 y_2 H_{\nu_3-1}^{(2)}(y_2) + (1-2 \nu_3) H_{\nu_3}^{(2)}(y_2) \Big) \notag \\
    & \quad \cdot \Big( \alpha_2 H_{\nu_2}^{(1)}(y_1) + \beta_2 H_{\nu_2}^{(2)}(y_1) \Big) \Big),
\end{align}
\begin{align}
    \beta_3 &= \frac{\pi }{8 y_1} \sqrt{-\frac{y_1}{y_2}} \Big(2 \alpha _2 y_1 y_2 H_{\nu _2-1}^{(1)}(y_1) H_{\nu _3}^{(1)}(y_2) \notag \\
    & \quad +\alpha _2 H_{\nu _2}^{(1)}(y_1) \Big( \big(2 \nu _3 y_1-2 \nu_2 y_2-y_1+y_2\big) H_{\nu _3}^{(1)}(y_2)-2 y_1 y_2 H_{\nu _3-1}^{(1)}(y_2) \Big) \notag \\
    & \quad -2 \beta _2 y_1 y_2 H_{\nu _3-1}^{(1)}(y_2) H_{\nu _2}^{(2)}(y_1) +\beta_2 H_{\nu _3}^{(1)}(y_2) \Big(2 y_1 y_2 H_{\nu _2-1}^{(2)}(y_1) \notag \\
    & \quad +\big(2 \nu _3 y_1-2 \nu _2 y_2-y_1+y_2\big) H_{\nu _2}^{(2)}(y_1) \Big) \Big),
\end{align}
\begin{align}
    \alpha_4 &= \frac{\pi }{8 z_1} \sqrt{-\frac{z_1}{z_2}} \Big(2 z_1 \Big(H_{\frac{3}{2}}^{(2)}(z_2)-z_2 H_{\frac{1}{2}}^{(2)}(z_2)\Big) \Big(\alpha _3 H_{\nu _3}^{(1)}(z_1)+\beta _3 H_{\nu
   _3}^{(2)}(z_1)\Big) \notag \\
   & \quad +z_2 H_{\frac{3}{2}}^{(2)}(z_2) \Big(2 \alpha _3 z_1 H_{\nu _3-1}^{(1)}(z_1)+\big(\alpha _3-2 \alpha _3 \nu _3\big) H_{\nu
   _3}^{(1)}(z_1) \notag \\
   & \quad +2 \beta _3 z_1 H_{\nu _3-1}^{(2)}(z_1)+\big(\beta _3-2 \beta _3 \nu _3\big) H_{\nu _3}^{(2)}(z_1)\Big)\Big),
\intertext{}
\begin{split}
    \beta_4 &=-\frac{\pi }{8 z_1} \sqrt{-\frac{z_1}{z_2}} \Big(H_{\frac{3}{2}}^{(1)}(z_2) \Big(2 \alpha _3 z_1 z_2 H_{\nu _3-1}^{(1)}(z_1) \\
    & \quad +\alpha _3 \big(-2 \nu _3 z_2+2 z_1+z_2\big) H_{\nu
   _3}^{(1)}(z_1) \\
   & \quad +2 \beta _3 z_1 z_2 H_{\nu _3-1}^{(2)}(z_1)+\beta _3 \big(-2 \nu _3 z_2+2 z_1+z_2\big) H_{\nu _3}^{(2)}(z_1)\Big) \\
   & \quad -2 z_1 z_2
   H_{\frac{1}{2}}^{(1)}(z_2) \Big(\alpha _3 H_{\nu _3}^{(1)}(z_1)+\beta _3 H_{\nu _3}^{(2)}(z_1)\Big)\Big),
   \end{split}
\end{align}
where $x_1=k(\tau_{R,1}-\tau_1)$, $x_2=k(\tau_{R,2}-\tau_1)$, $y_1=k(\tau_{R,2}-\tau_2)$, $y_2=k(\tau_{R,3}-\tau_2)$, $z_1=k(\tau_{R,3}-\tau_3)$, $z_2=k(\tau_{R,4}-\tau_3)$.
Since the results are too complex, we only provide the recursive formulas for $\alpha_j$ and $\beta_j$.

\begin{figure}[htbp]
    \centering
\subfigure[ ]
{\includegraphics[width=0.46\textwidth]{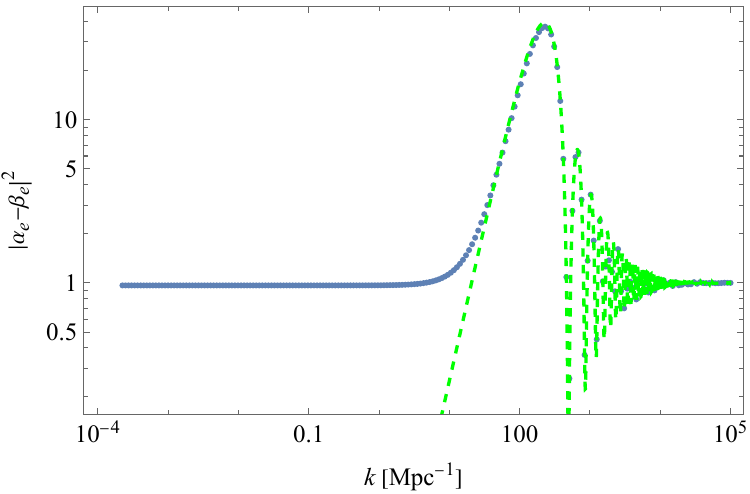}\label{fig:jinsi}}
    \quad
\subfigure[ ]
{\includegraphics[width=0.47\textwidth]{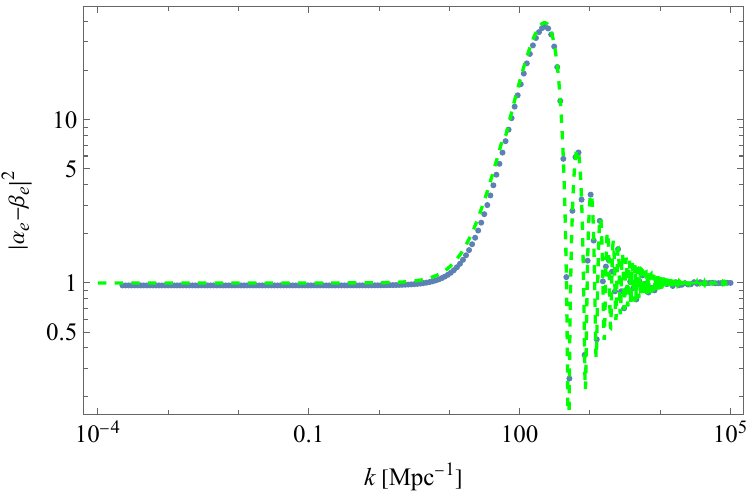}\label{fig:jinsi2}}
\caption{Left: Comparison between $|G_1|^2$ (green dashed curve) and the numerical result of ${P_T/ P_{T,inf2}}$ (blue dotted curve). Right: Comparison between the template (\ref{eq:temp01}) (green dashed curve) and the numerical result of ${P_T/ P_{T,inf2}}$ (blue dotted curve). We have set $\bar{\cal H}_1=10^4a_0H_0$, where $a_0=1$ and $H_0=67\, {\rm km/s/Mpc}$. We also set $w_2=-15$ and $w_3={2}/{3}$, which corresponding to $\nu_2={6}/{11}$ and $\nu_3={1}/{6}$, respectively.} \label{phidotphi}
\end{figure}

Next, we attempt to explore potential approximations. It's worth noting that in our model, $y_2$ and $z_1$ are negative. Therefore, our approach here is to consider $k \gg \bar{\mathcal{H}}_1$. By series expansion, we get
\begin{align}
    G_1 &={\alpha_4-\beta_4}\\
    &=\frac{\pi ^{3/2}}{16 \sqrt{2} z_2^2} \bigg(2 H_{\nu _2-1}^{(1)}\left(y_1\right) \bigg(H_{\nu _3}^{(1)}\left(y_2\right) \bigg(2 H_{\nu _3-1}^{(2)}\left(z_1\right) \sqrt{y_1 y_2 z_1} z_2 \left(\cos \left(z_2\right) z_2-\sin
   \left(z_2\right)\right)\notag\\
   &+H_{\nu _3}^{(2)}\left(z_1\right) \bigg(\sin \left(z_2\right) \left(2 \sqrt{y_1 y_2 z_1} z_2^2+\sqrt{\frac{y_1 y_2}{z_1}} \left(1-2 \nu _3\right) z_2-2 \sqrt{y_1 y_2
   z_1}\right)\notag\\
   &+\cos \left(z_2\right) z_2 \left(\sqrt{\frac{y_1 y_2}{z_1}} z_2 \left(2 \nu _3-1\right)+2 \sqrt{y_1 y_2 z_1}\right)\bigg)\bigg)\notag\\
   &+H_{\nu _3}^{(2)}\left(y_2\right) \bigg(H_{\nu
   _3}^{(1)}\left(z_1\right) \bigg(\sin \left(z_2\right) \left(-2 \sqrt{y_1 y_2 z_1} z_2^2+\sqrt{\frac{y_1 y_2}{z_1}} \left(2 \nu _3-1\right) z_2+2 \sqrt{y_1 y_2 z_1}\right)\notag\\
    &-\cos \left(z_2\right) z_2
   \left(\sqrt{\frac{y_1 y_2}{z_1}} z_2 \left(2 \nu _3-1\right)+2 \sqrt{y_1 y_2 z_1}\right)\bigg)-2 H_{\nu _3-1}^{(1)}\left(z_1\right) \sqrt{y_1 y_2 z_1} z_2 \big(\cos \left(z_2\right) z_2\notag\\
   &-\sin
   \left(z_2\right)\big)\bigg)\bigg)+H_{\nu _2}^{(1)}\left(y_1\right) \bigg(-4 \cos \left(z_2\right) H_{\nu _3}^{(1)}\left(y_2\right) H_{\nu _3}^{(2)}\left(z_1\right) \sqrt{\frac{y_1}{y_2 z_1}} \nu
   _3^2 z_2^2\notag\\
   &+2 \cos \left(z_2\right) H_{\nu _3}^{(1)}\left(y_2\right) H_{\nu _3}^{(2)}\left(z_1\right) \sqrt{\frac{y_2}{y_1 z_1}} \nu _2 z_2^2-4 \cos \left(z_2\right) H_{\nu _3}^{(1)}\left(y_2\right)
   H_{\nu _3-1}^{(2)}\left(z_1\right) \sqrt{\frac{y_2 z_1}{y_1}} \nu _2 z_2^2\notag\\
   &-4 H_{\nu _3}^{(1)}\left(y_2\right) H_{\nu _3}^{(2)}\left(z_1\right) \sin \left(z_2\right) \sqrt{\frac{y_2 z_1}{y_1}} \nu _2
   z_2^2-4 \cos \left(z_2\right) H_{\nu _3}^{(1)}\left(y_2\right) H_{\nu _3}^{(2)}\left(z_1\right) \sqrt{\frac{y_2}{y_1 z_1}} \nu _2 \nu _3 z_2^2\notag\\
   &+4 \cos \left(z_2\right) H_{\nu _3}^{(1)}\left(y_2\right)
   H_{\nu _3}^{(2)}\left(z_1\right) \sqrt{\frac{y_1}{y_2 z_1}} \nu _3 z_2^2+2 \cos \left(z_2\right) H_{\nu _3}^{(1)}\left(y_2\right) H_{\nu _3}^{(2)}\left(z_1\right) \sqrt{\frac{y_2}{y_1 z_1}} \nu _3
   z_2^2\notag\\
   &-4 \cos \left(z_2\right) H_{\nu _3-1}^{(1)}\left(y_2\right) H_{\nu _3}^{(2)}\left(z_1\right) \sqrt{\frac{y_1 y_2}{z_1}} \nu _3 z_2^2-4 \cos \left(z_2\right) H_{\nu _3}^{(1)}\left(y_2\right) H_{\nu
   _3-1}^{(2)}\left(z_1\right) \sqrt{\frac{y_1 z_1}{y_2}} \nu _3 z_2^2\notag\\
   &-4 H_{\nu _3}^{(1)}\left(y_2\right) H_{\nu _3}^{(2)}\left(z_1\right) \sin \left(z_2\right) \sqrt{\frac{y_1 z_1}{y_2}} \nu _3
   z_2^2-\cos \left(z_2\right) H_{\nu _3}^{(1)}\left(y_2\right) H_{\nu _3}^{(2)}\left(z_1\right) \sqrt{\frac{y_1}{y_2 z_1}} z_2^2\notag\\
   &-\cos \left(z_2\right) H_{\nu _3}^{(1)}\left(y_2\right) H_{\nu
   _3}^{(2)}\left(z_1\right) \sqrt{\frac{y_2}{y_1 z_1}} z_2^2+2 \cos \left(z_2\right) H_{\nu _3-1}^{(1)}\left(y_2\right) H_{\nu _3}^{(2)}\left(z_1\right) \sqrt{\frac{y_1 y_2}{z_1}} z_2^2\notag\\
   &+2 \cos
   \left(z_2\right) H_{\nu _3}^{(1)}\left(y_2\right) H_{\nu _3-1}^{(2)}\left(z_1\right) \sqrt{\frac{y_1 z_1}{y_2}} z_2^2+2 H_{\nu _3}^{(1)}\left(y_2\right) H_{\nu _3}^{(2)}\left(z_1\right) \sin
   \left(z_2\right) \sqrt{\frac{y_1 z_1}{y_2}} z_2^2\notag\\
   &+2 \cos \left(z_2\right) H_{\nu _3}^{(1)}\left(y_2\right) H_{\nu _3-1}^{(2)}\left(z_1\right) \sqrt{\frac{y_2 z_1}{y_1}} z_2^2+2 H_{\nu
   _3}^{(1)}\left(y_2\right) H_{\nu _3}^{(2)}\left(z_1\right) \sin \left(z_2\right) \sqrt{\frac{y_2 z_1}{y_1}} z_2^2\notag\\
   &-4 \cos \left(z_2\right) H_{\nu _3-1}^{(1)}\left(y_2\right) H_{\nu
   _3-1}^{(2)}\left(z_1\right) \sqrt{y_1 y_2 z_1} z_2^2-4 H_{\nu _3-1}^{(1)}\left(y_2\right) H_{\nu _3}^{(2)}\left(z_1\right) \sin \left(z_2\right) \sqrt{y_1 y_2 z_1} z_2^2\notag\\
   &+4 H_{\nu
   _3}^{(1)}\left(y_2\right) H_{\nu _3}^{(2)}\left(z_1\right) \sin \left(z_2\right) \sqrt{\frac{y_1}{y_2 z_1}} \nu _3^2 z_2-2 H_{\nu _3}^{(1)}\left(y_2\right) H_{\nu _3}^{(2)}\left(z_1\right) \sin
   \left(z_2\right) \sqrt{\frac{y_2}{y_1 z_1}} \nu _2 z_2\notag\\
   &-4 \cos \left(z_2\right) H_{\nu _3}^{(1)}\left(y_2\right) H_{\nu _3}^{(2)}\left(z_1\right) \sqrt{\frac{y_2 z_1}{y_1}} \nu _2 z_2+4 H_{\nu
   _3}^{(1)}\left(y_2\right) H_{\nu _3-1}^{(2)}\left(z_1\right) \sin \left(z_2\right) \sqrt{\frac{y_2 z_1}{y_1}} \nu _2 z_2\notag\\
   &+4 H_{\nu _3}^{(1)}\left(y_2\right) H_{\nu _3}^{(2)}\left(z_1\right) \sin
   \left(z_2\right) \sqrt{\frac{y_2}{y_1 z_1}} \nu _2 \nu _3 z_2-4 H_{\nu _3}^{(1)}\left(y_2\right) H_{\nu _3}^{(2)}\left(z_1\right) \sin \left(z_2\right) \sqrt{\frac{y_1}{y_2 z_1}} \nu _3 z_2\notag\\
   &-2 H_{\nu
   _3}^{(1)}\left(y_2\right) H_{\nu _3}^{(2)}\left(z_1\right) \sin \left(z_2\right) \sqrt{\frac{y_2}{y_1 z_1}} \nu _3 z_2+4 H_{\nu _3-1}^{(1)}\left(y_2\right) H_{\nu _3}^{(2)}\left(z_1\right) \sin
   \left(z_2\right) \sqrt{\frac{y_1 y_2}{z_1}} \nu _3 z_2\notag
    \intertext{}
\begin{split}
    &-4 \cos \left(z_2\right) H_{\nu _3}^{(1)}\left(y_2\right) H_{\nu _3}^{(2)}\left(z_1\right) \sqrt{\frac{y_1 z_1}{y_2}} \nu _3 z_2+4 H_{\nu
   _3}^{(1)}\left(y_2\right) H_{\nu _3-1}^{(2)}\left(z_1\right) \sin \left(z_2\right) \sqrt{\frac{y_1 z_1}{y_2}} \nu _3 z_2\notag\\
    &+2 H_{\nu _3-1}^{(1)}\left(z_1\right) \left(\cos \left(z_2\right) z_2-\sin \left(z_2\right)\right) \bigg(2 \sqrt{y_1 y_2 z_1} H_{\nu _3-1}^{(2)}\left(y_2\right)+H_{\nu _3}^{(2)}\left(y_2\right) \bigg(2 \sqrt{\frac{y_2 z_1}{y_1}} \nu _2\notag\\
    &+2 \sqrt{\frac{y_1 z_1}{y_2}} \nu _3-\sqrt{\frac{y_1 z_1}{y_2}}-\sqrt{\frac{y_2 z_1}{y_1}}\bigg)\bigg) z_2+H_{\nu _3}^{(1)}\left(y_2\right) H_{\nu _3}^{(2)}\left(z_1\right) \sin \left(z_2\right) \sqrt{\frac{y_1}{y_2 z_1}} z_2\notag\\
    &+H_{\nu _3}^{(1)}\left(y_2\right) H_{\nu _3}^{(2)}\left(z_1\right) \sin \left(z_2\right) \sqrt{\frac{y_2}{y_1 z_1}} z_2-2 H_{\nu _3-1}^{(1)}\left(y_2\right) H_{\nu _3}^{(2)}\left(z_1\right) \sin \left(z_2\right) \sqrt{\frac{y_1 y_2}{z_1}} z_2\notag\\
    &+2 \cos \left(z_2\right) H_{\nu _3}^{(1)}\left(y_2\right) H_{\nu _3}^{(2)}\left(z_1\right) \sqrt{\frac{y_1 z_1}{y_2}} z_2-2 H_{\nu _3}^{(1)}\left(y_2\right) H_{\nu _3-1}^{(2)}\left(z_1\right) \sin \left(z_2\right) \sqrt{\frac{y_1 z_1}{y_2}} z_2\notag\\
    &+2 \cos \left(z_2\right) H_{\nu _3}^{(1)}\left(y_2\right) H_{\nu _3}^{(2)}\left(z_1\right) \sqrt{\frac{y_2 z_1}{y_1}} z_2-2 H_{\nu _3}^{(1)}\left(y_2\right) H_{\nu _3-1}^{(2)}\left(z_1\right) \sin \left(z_2\right) \sqrt{\frac{y_2 z_1}{y_1}} z_2\notag\\
    &-4 \cos \left(z_2\right) H_{\nu _3-1}^{(1)}\left(y_2\right) H_{\nu _3}^{(2)}\left(z_1\right) \sqrt{y_1 y_2 z_1} z_2+4 H_{\nu _3-1}^{(1)}\left(y_2\right) H_{\nu _3-1}^{(2)}\left(z_1\right) \sin \left(z_2\right) \sqrt{y_1 y_2 z_1} z_2\notag\\
    &+4 H_{\nu _3}^{(1)}\left(y_2\right) H_{\nu _3}^{(2)}\left(z_1\right) \sin \left(z_2\right) \sqrt{\frac{y_2 z_1}{y_1}} \nu _2+4 H_{\nu _3}^{(1)}\left(y_2\right) H_{\nu _3}^{(2)}\left(z_1\right) \sin \left(z_2\right) \sqrt{\frac{y_1 z_1}{y_2}} \nu _3\notag\\
    &+H_{\nu _3}^{(1)}\left(z_1\right) \bigg(2 H_{\nu _3-1}^{(2)}\left(y_2\right) \bigg(\sin \left(z_2\right) \bigg(2 \sqrt{y_1 y_2 z_1} z_2^2+\sqrt{\frac{y_1 y_2}{z_1}} \left(1-2 \nu _3\right) z_2-2 \sqrt{y_1 y_2 z_1}\bigg)\notag\\
    &+\cos \left(z_2\right) z_2 \left(\sqrt{\frac{y_1 y_2}{z_1}} z_2 \left(2 \nu _3-1\right)+2 \sqrt{y_1 y_2 z_1}\right)\bigg)+H_{\nu _3}^{(2)}\left(y_2\right) \bigg(\cos \left(z_2\right) z_2 \bigg(4 \sqrt{\frac{y_2 z_1}{y_1}} \nu _2\notag\\
    &+4 \sqrt{\frac{y_1 z_1}{y_2}} \nu _3+z_2 \left(2 \nu _3-1\right) \left(2 \sqrt{\frac{y_2}{y_1 z_1}} \nu _2+2 \sqrt{\frac{y_1}{y_2 z_1}} \nu _3-\sqrt{\frac{y_1}{y_2 z_1}}-\sqrt{\frac{y_2}{y_1 z_1}}\right)-2 \sqrt{\frac{y_1 z_1}{y_2}}\notag\\
    &-2 \sqrt{\frac{y_2 z_1}{y_1}}\bigg)+\sin \left(z_2\right) \bigg(\bigg(-2 \bigg(\sqrt{\frac{y_1 z_1}{y_2}}+\sqrt{\frac{y_2 z_1}{y_1}}\bigg)+4 \sqrt{\frac{y_2 z_1}{y_1}} \nu _2+4 \sqrt{\frac{y_1 z_1}{y_2}} \nu _3\bigg)z_2^2\notag\\
    &-\left(2 \nu _3-1\right) \left(2 \sqrt{\frac{y_2}{y_1 z_1}} \nu _2+2 \sqrt{\frac{y_1}{y_2 z_1}} \nu _3-\sqrt{\frac{y_1}{y_2 z_1}}-\sqrt{\frac{y_2}{y_1 z_1}}\right) z_2+2 \bigg(-2 \sqrt{\frac{y_2 z_1}{y_1}} \nu _2\notag\\
    &-2 \sqrt{\frac{y_1 z_1}{y_2}} \nu _3+\sqrt{\frac{y_1 z_1}{y_2}}+\sqrt{\frac{y_2 z_1}{y_1}}\bigg)\bigg)\bigg)\bigg)-2 H_{\nu _3}^{(1)}\left(y_2\right) H_{\nu _3}^{(2)}\left(z_1\right) \sin \left(z_2\right) \sqrt{\frac{y_1 z_1}{y_2}}\notag\\
    &-2 H_{\nu _3}^{(1)}\left(y_2\right) H_{\nu _3}^{(2)}\left(z_1\right) \sin \left(z_2\right) \sqrt{\frac{y_2 z_1}{y_1}}+4 H_{\nu _3-1}^{(1)}\left(y_2\right) H_{\nu _3}^{(2)}\left(z_1\right) \sin \left(z_2\right) \sqrt{y_1 y_2 z_1}\bigg)\bigg).
\end{split}
\end{align}
This expression remains quite complex, although it is much simpler compared to the original expression for $\alpha_4-\beta_4$.
We plot $|\alpha_4-\beta_4|^2=|G_1|^2$ in Fig. \ref{fig:jinsi}. It can be observed that, in the limit where $k \gg \bar{\mathcal{H}}_1$, this approximation matches the numerical solution quite well.

To better match the numerical results across the entire frequency range, we propose a parameterized template as
\begin{equation}
    {P_T(k)\over P_{T,inf2}} = \frac{1}{1+e^{k-{\bar{\mathcal{H}}}_3}}+|G_1|^2 \,. \label{eq:temp01}
\end{equation}
We compare this template with the numerical results in Fig. \ref{fig:jinsi2}. It is evident that Eq. (\ref{eq:temp01}) serves as a reliable approximation for observational purposes.



\bibliographystyle{utphys}

\bibliography{240822Ref}


 \end{document}